\documentclass[journal]{IEEEtran}

\usepackage{amsmath,amssymb,mathtools}
\usepackage{algorithm}
\usepackage{algpseudocode}
\usepackage{graphicx}
\usepackage{cite}
\usepackage{booktabs}
\usepackage{url}
\usepackage{comment}
\usepackage{subcaption}

\newcommand{\E}{\mathcal{E}}
\newcommand{\X}{\mathbf{x}}
\newcommand{\U}{\mathbf{u}}
\newcommand{\R}{\mathbb{R}}
\newcommand{\softplus}{\mathrm{softplus}}
\newcommand{\ODESolve}{\mathrm{ODESolve}}
\newcommand{\abs}[1]{\left|#1\right|}

\begin{document}

\title{Receding-Horizon Maximum-Likelihood Estimation of Neural-ODE Dynamics and Thresholds from Event Cameras}
\author{Kazumune Hashimoto, Kazunobu Serizawa, Masako Kishida \thanks{Kazumune Hashimoto and Kazunobu Serizawa are with the Graduate School of Engineering, The University of Osaka, Suita, Japan (e-mail: hashimoto@eei.eng.osaka-u.ac.jp, serizawa.kazunobu.ogp@ecs.osaka-u.ac.jp). Masako Kishida is with the Faculty of Engineering, Information and Systems, University of Tsukuba (e-mail: kishida@ieee.org).}
}
\maketitle

\begin{abstract}
Event cameras produce asynchronous brightness-change events when the accumulated log-intensity change at a pixel since its previous event reaches a threshold, yielding a history-dependent observation process. We study online maximum-likelihood identification of continuous-time dynamics directly from such raw event streams. The latent state evolves according to a Neural ODE and is mapped to predicted log-intensity through a differentiable state-to-image model. We model the event stream as a history-dependent marked temporal point process whose conditional intensity is a smooth probabilistic surrogate of contrast-threshold triggering, and we treat the pixel-dependent contrast thresholds as unknown parameters to be estimated jointly with the dynamics. This yields a normalized likelihood over event times and marks consisting of an event term and a compensator integral. To enable streaming inference, we develop a fixed-lag receding-horizon estimator with compact per-pixel boundary memory and Monte Carlo pixel subsampling for the compensator. Synthetic experiments demonstrate joint recovery of the dynamics parameters and the contrast threshold map, and quantify the resulting accuracy--latency trade-off as the horizon length varies.
\end{abstract}


\section{Introduction}
Event cameras~\cite{gallego2022survey}, including the Dynamic Vision Sensor (DVS)~\cite{lichtsteiner2008dvs} and the DAVIS family~\cite{brandli2014davis}, report asynchronous brightness-change events with microsecond timestamps, high dynamic range, and low latency.
Unlike frame-based sensors, they transmit information only when brightness at each pixel changes sufficiently, thereby emphasizing moving edges and reducing motion blur caused by exposure time integration.
These properties make event streams well-suited to high-speed and high-dynamic-range settings, enabling optical flow~\cite{zhu2018evflownet,gehrig2021eraft,gehrig2024denseflow}; high-frame-rate video reconstruction and frame interpolation~\cite{rebecq2019eventstovideo,scheerlinck2018intensity,pan2022hfr,tulyakov2021timelens}; and ego-motion/SLAM datasets and simulators~\cite{mueggler2017dataset,rebecq2018esim}.
Recent work further highlights event-driven sensing and event-camera processing for efficient perception, video reconstruction, and stereo depth estimation~\cite{liu2019eventdriven,gu2024adaptiveparam,chen2024eventstereodepth}.

Many event-based pipelines target a single-shot prediction from a fixed spatiotemporal event representation.
In contrast, a broad class of applications requires continuous-time state and dynamics estimates for tracking, prediction, online system identification, and feedback control.
In these settings, the information needed for estimation is encoded not only in the event marks (pixel and polarity) but also in fine event timing.
This motivates estimators that (i) respect asynchronous timestamps and (ii) provide a likelihood over the raw event stream, so that dynamics parameters can be fit and adapted online in a principled way.
Asynchronous thresholded observations have been studied in time encoding, level-crossing sampling, unlimited sampling, and spike-train reconstruction, where information is represented by event times rather than uniform samples~\cite{lazar2004perfect,senay2012levelcrossing,cancimino2014asynchronous,florescu2022timeencoding,chattopadhyay2025spiketrain}.
In parallel, statistical modeling and inference for event times has continued to develop in temporal point-process and event-triggered estimation settings~\cite{snyder1972doublypoisson,smith2003ssmpp,eden2004ppadaptive,zhou2024epogda,liu2025confidence,rong2025ppclass,dong2025tppmix}.
Event-camera streams combine both aspects: each pixel generates asynchronous thresholded observations, while the full sensor output is naturally a marked event sequence whose information is carried jointly by event times and marks.

A common description of event generation is contrast thresholding: a pixel fires when the accumulated change in log-intensity since its previous event reaches a threshold.
Because the threshold directly affects event timing, it directly impacts dynamics estimation from timestamps.
In practice, the effective threshold is often not known precisely and can depend on sensor settings and operating conditions~\cite{stoffregen2020sim2real,wang2020contrastcalib}.
Treating the threshold as a fixed \emph{known} constant can therefore introduce bias, especially when the inference relies on fine timing structure.
Beyond threshold uncertainty, event streams pose fundamental modeling challenges.
Event times are irregular, observations are discrete marks (pixel location and polarity), and event generation is history-dependent because each pixel resets its reference after it fires.
A common workaround is to aggregate events into fixed-rate tensors (for example, voxel grids) and apply standard architectures~\cite{zhu2018evflownet,rebecq2019eventstovideo,gu2024adaptiveparam,chen2024eventstereodepth,zubic2024ssm}.
Such representations are effective for downstream prediction but discretize time and typically do not define a normalized likelihood over the original timestamps.
Complementary model-based approaches incorporate event-generation mechanisms for reconstruction and filtering, or optimize surrogate objectives such as contrast maximization and alignment under a motion hypothesis~\cite{scheerlinck2018intensity,gallego2018contrastmax,stoffregen2019contrastmax,gu2021eventalign}.
However, these approaches are usually not formulated as maximum-likelihood estimation of continuous-time dynamics directly from the raw event stream.

\begin{figure}[t]
  \centering
  \includegraphics[width=0.9\linewidth]{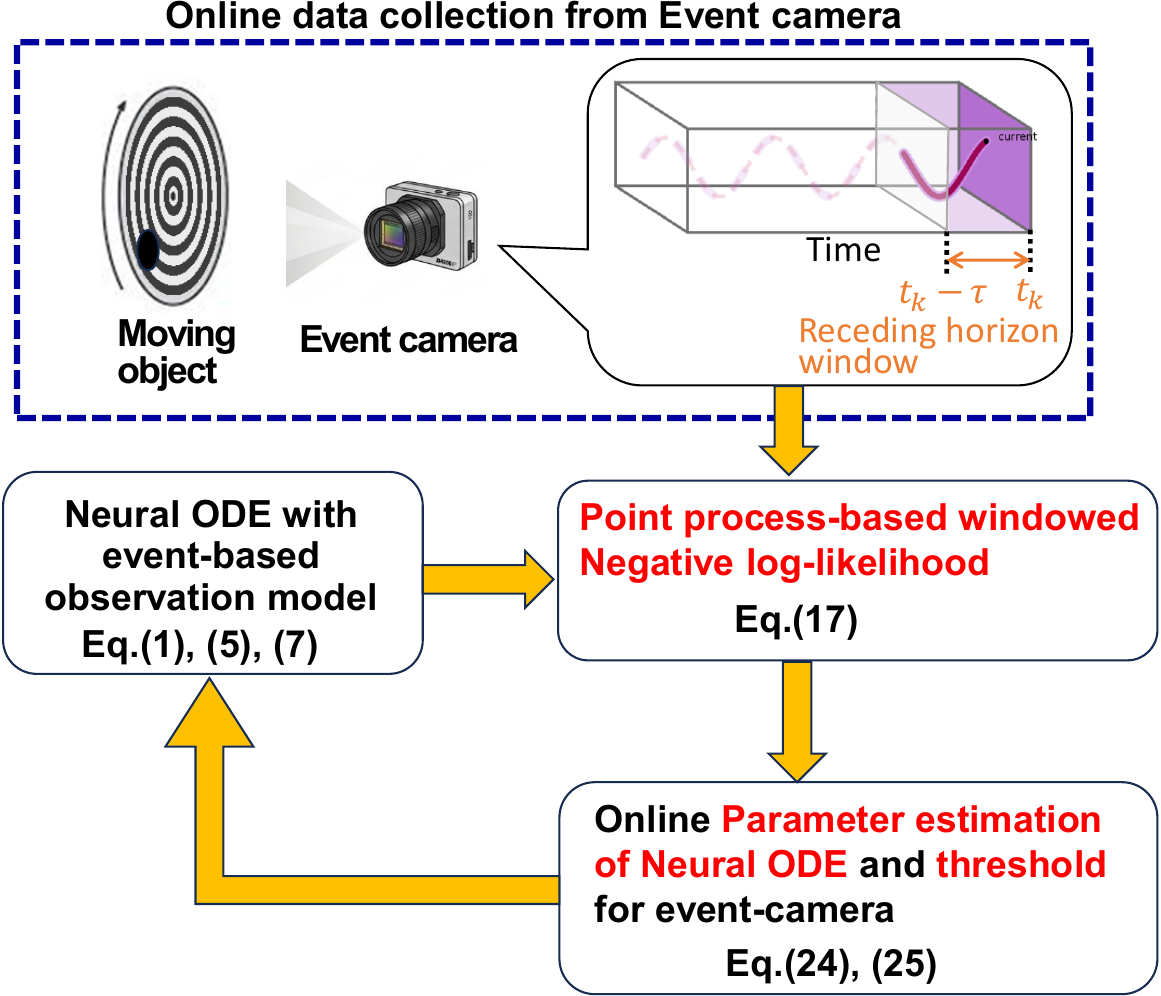}
  \caption{Illustration of the proposed framework.}
  \label{fig:proposedillust}
\end{figure}

A likelihood-based alternative is to model an event stream as a marked temporal point process.
This provides a normalized probability model over both event times and marks, and it naturally accommodates irregular timestamps and history dependence.
A practical challenge is computational cost.
Point-process likelihoods include, in addition to a sum over observed events, a normalization (no-event) term that integrates the predicted event rate over time and sums over the full pixel grid.
Evaluating this term can dominate runtime especially for long streams, which is particularly problematic in online estimation settings.
Sliding-window and moving/receding-horizon estimators, as well as recent event-triggered online estimation methods, provide a natural route to bounded-latency updates in dynamical systems~\cite{rao2001mhe,rao2003nonlinearmhe,battistelli2017binarymhe,shi2014eventtriggered,zou2020mhereview,zhou2024epogda,liu2025confidence}.
For event cameras, however, the observation model is both high-dimensional and per-pixel history dependent, which makes a direct application of existing methods nontrivial.

In this paper, we develop an online maximum-likelihood estimator for continuous-time dynamics observed through event-camera streams.
Its latent state follows a Neural ODE~\cite{chen2018neuralode} and is mapped to predicted log-intensity through a differentiable state-to-image model.
Events are modeled with a history-dependent marked point process whose conditional intensity is a smooth surrogate of contrast-threshold triggering.
We treat the pixel-dependent contrast thresholds as unknown parameters and estimate them jointly with the dynamics.
To support streaming operation, we maintain a compact per-pixel memory that summarizes the information needed to evaluate history-dependent residuals.
We use a receding-horizon scheme that replays only recent events, aiming to keep per-update computation bounded.

The contributions of this paper are twofold:
\begin{itemize}
\item
We introduce a differentiable residual-to-rate mapping as a smooth surrogate of contrast-threshold triggering within a \emph{marked point-process likelihood}, enabling joint estimation of dynamics parameters and pixel-dependent contrast thresholds.
\item
We propose a \emph{receding-horizon update} that performs a few replayed gradient steps per update on a fixed-lag window, alleviating the computational cost compared with offline fitting.
\end{itemize}
We evaluate the estimator on a synthetic sequence and report ablations that characterize accuracy--latency trade-offs, including recovery of dynamics parameters and the contrast threshold.

\textit{Related work:}
Our work connects four lines of research.

\emph{Event-driven sensing and asynchronous threshold acquisition.}
Recent work studies event-driven sensing and event-camera processing for efficient perception, video reconstruction, and stereo depth estimation~\cite{liu2019eventdriven,gu2024adaptiveparam,chen2024eventstereodepth}.
Closely related time-encoding and level-crossing/unlimited-sampling work studies how continuous-time signals can be represented and recovered from threshold-crossing times~\cite{lazar2004perfect,senay2012levelcrossing,cancimino2014asynchronous,florescu2022timeencoding,chattopadhyay2025spiketrain}.
Event cameras can be viewed as a spatially distributed, polarity-marked extension of this principle.
Unlike classical time-encoding setups, however, the quantity of interest here is not the per-pixel brightness trace itself but latent continuous-time dynamics observed through a state-to-image map, and the thresholds are not assumed known.

\emph{Event-camera pipelines and model-based objectives.}
Many approaches aggregate events into fixed-rate tensors (e.g., voxel grids) and apply standard deep architectures for tasks such as optical flow and video reconstruction~\cite{zhu2018evflownet,rebecq2019eventstovideo,gu2024adaptiveparam,chen2024eventstereodepth}.
These methods are effective for downstream prediction.
However, the aggregation discretizes time and typically does not yield a \emph{normalized likelihood} over the raw event timestamps.
Complementary approaches incorporate event-generation mechanisms for reconstruction and filtering, or optimize surrogate objectives such as contrast maximization and alignment under a motion hypothesis~\cite{scheerlinck2018intensity,gallego2018contrastmax,stoffregen2019contrastmax,gu2021eventalign}.
In particular, the spatio-temporal Poisson point-process model in~\cite{gu2021eventalign} is closer in spirit because it defines a likelihood over event data, but it is aimed at event alignment under a motion hypothesis rather than joint estimation of continuous-time latent dynamics and pixel-dependent thresholds from a per-pixel history-dependent observation model.

\emph{Statistical inference with point-process or event-triggered observations.}
Classical work on counting/point-process observations studies filtering and state-space inference for latent dynamical systems observed through event times~\cite{snyder1972doublypoisson,segall1975counting,smith2003ssmpp,eden2004ppadaptive,zammitmangion2011onlinepp}.
Recent work further studies point-process learning, classification, and event-triggered online estimation~\cite{rong2025ppclass,dong2025tppmix,zhou2024epogda,liu2025confidence}.
In a related vein, moving-horizon estimation with binary or threshold sensors has been developed for lower-dimensional dynamical systems~\cite{rao2001mhe,rao2003nonlinearmhe,battistelli2017binarymhe,shi2014eventtriggered,zou2020mhereview}.
Our setting differs in that the mark space is extremely large (pixel location and polarity), the observation model is per-pixel history dependent due to the reset at each firing event, and the thresholds are learned jointly with the dynamics.

\emph{Continuous-time neural dynamics.}
Neural temporal point-process models such as RMTPP~\cite{du2016rmtpp} and Neural Hawkes~\cite{mei2017neuralhawkes} provide flexible likelihoods for marked sequences, while continuous-time neural models such as Neural ODEs and latent/jump variants provide convenient representations of dynamics under irregular observations~\cite{chen2018neuralode,rubanova2019latentode,herrera2020njode}.
Our work exploits the event-camera structure explicitly by (i) defining the intensity through a contrast-threshold residual computed from a Neural ODE state via a differentiable state-to-image model, and (ii) making online maximum-likelihood updates practical using a compact per-pixel memory and a fixed-lag receding-horizon scheme with bounded per-update computation.

\section{Background}

\subsection{Neural ODEs for continuous-time dynamics}
A Neural Ordinary Differential Equation (Neural ODE) represents latent dynamics in continuous time by learning a vector field.
Let $\X(t)\in\R^n$ denote the latent state at continuous time $t$.
A Neural ODE specifies the initial value problem: 
\begin{equation}\label{eq:neuralode}
\frac{d\X(t)}{dt}=f_{\vartheta}(\X(t),t),
\end{equation}
where $f_{\vartheta}$ is parameterized by trainable parameters $\vartheta$ (e.g., neural-network weights and/or physical parameters).
Given an initial condition $\X(t_0)$, the state at time $t$ satisfies
\begin{equation}\label{eq:neuralode_solution}
\X(t)=\X(t_0)+\int_{t_0}^{t} f_{\vartheta}(\X(\tau),\tau)\,d\tau.
\end{equation}
We obtain $\X(t)$ numerically with an ODE solver, which we denote by 
\[
\X(t)=\ODESolve \big(f_{\vartheta},\X(t_0),t_0,t\big).
\] 
The ODE parameters are learned by minimizing a given task loss
\[
\min_{\vartheta}\ {\ell}(\vartheta),
\]
where ${\ell}$ is typically defined from data-fit terms. Training requires gradients of ${\ell}$ with respect to $\vartheta$.
Backpropagating through the internal steps of an ODE solver can be memory intensive for long time horizons. Neural ODEs are therefore often trained using the adjoint sensitivity method~\cite{chen2018neuralode}.
This introduces the adjoint state
\begin{equation}\label{eq:adjoint_def}
\mathbf{a}(t) = \frac{\partial {\ell}(\vartheta)}{\partial \mathbf{x}(t)},
\end{equation}
which propagates sensitivities backward in time and yields parameter gradients via \(\partial f_{\vartheta}/\partial \vartheta\).

\subsection{Event cameras and contrast-threshold events}
An event camera outputs an asynchronous stream of per-pixel brightness-change events
\begin{equation}\label{eq:event_stream}
\E=\{(\U_k,t_k,p_k)\}_{k=1}^{K},
\end{equation}
where $\U_k\in\Omega=\{0,\dots,W_{\mathrm{img}}-1\}\times\{0,\dots,H_{\mathrm{img}}-1\}$ is a 2-D pixel location,
$t_k\in\R_{\ge 0}$ is a timestamp, and $p_k\in\{+1,-1\}$ is the polarity (sign of the brightness change).

A common generative description is based on \emph{log-intensity}
$L(\U,t)=\log(I(\U,t)+\varepsilon)$ (with small $\varepsilon>0$ for numerical stability) for any pixel location $\U\in\Omega$ and an idealized contrast-threshold mechanism:
for each pixel, the sensor integrates the log-intensity change since the most recent event at that pixel and triggers a new event when the accumulated change reaches a threshold.
Let $t^{-}(\U;t)$ be the most recent event time at pixel $\U$ strictly before $t$.
Then, an event with polarity $p$ occurs at the \emph{first} time $t$ after $t^{-}(\U;t)$ such that
\begin{equation}\label{eq:event_rule}
L(\U,t)-L(\U,t^{-}(\U;t))=p\,C(\U),
\end{equation}
where $C(\U)>0$ is the (possibly pixel-dependent) contrast threshold. Two aspects are important for modeling and learning: \textbf{Per-pixel history dependence.}
The reference time $t^{-}(\U;t)$ resets after each event, so the residual that determines the next trigger depends on the last event time at the {same} pixel. \textbf{Unknown thresholds.}
Thresholds can vary across pixels and conditions.
For example, bias settings, temperature changes, sensor aging, or internal auto-calibration can shift the effective threshold over time.
Hence, $C(\U)$ is often not precisely known a priori (and may not be provided by the device API).
%Treating $C(\U)$ (or a compact parameterization thereof, e.g. $C(\U)=\exp(c_0+c_\U)$ or a low-rank field) as a learnable parameter can reduce systematic mismatch when fitting dynamics from event timing.

\subsection{Temporal marked point processes and conditional intensity}\label{sec:pp_background}
A temporal point process models random events occurring at continuous times.
A \emph{marked} point process associates a \emph{mark} $m$ to each event time.
For event cameras, a natural choice is to treat the mark as the pair
\[
m=(\U,p)\in\mathcal{M}=\Omega\times\{+1,-1\},
\]
i.e., the mark encodes \emph{which pixel fired} and \emph{with which polarity}.
In other words, this can be seen as a $(2W_{\mathrm{img}}H_{\mathrm{img}})$-dimensional multivariate point process, one dimension per $(\U,p)$.

Let $\E_{<t}$ denote the event history strictly before time $t$.
The \emph{conditional intensity} $\lambda_m(t\mid \E_{<t})$ is the instantaneous rate of observing an event with mark $m$ at time $t$:
{ 
\begin{equation*}
\begin{aligned}
\lambda_{m}(t\mid \E_{<t})
&=\lim_{\Delta t\rightarrow 0}\frac{1}{\Delta t}\\
&\quad\times \Pr\!\Big(\text{event with mark } m \text{ in }[t,t+\Delta t)\mid \E_{<t}\Big).
\end{aligned}
\end{equation*}
}
The corresponding total intensity (summing over all marks) is
\[
\Lambda(t\mid \E_{<t})=\sum_{m\in\mathcal{M}}\lambda_{m}(t\mid \E_{<t}).
\]

Given an observed event stream $\E=\{(\U_k,t_k,p_k)\}_{k=1}^{K}$ on $[t_0,T]$, define $m_k=(\U_k,p_k)$.
The log-likelihood of a marked point process has the standard form~\cite{daley2007pp}
\begin{equation}\label{eq:pp_ll_bg}
\log p(\E)=\sum_{k=1}^{K}\log\lambda_{m_k}(t_k\mid \E_{<t_k})
-\int_{t_0}^{T}\Lambda(t\mid \E_{<t})\,dt.
\end{equation}
The integral term is called the \emph{compensator} (also known as the \textit{survival term}).
It ensures proper normalization by accounting for the probability that no additional events occur beyond those observed.
Intuitively, it uses information from time and mark regions where the stream is silent, and it prevents degenerate solutions
in which the likelihood is increased by driving intensities to arbitrarily large values.

This formulation aligns with event-camera data in two ways.
First, event cameras produce irregular (or, non periodic) timestamps and discrete marks, and the intensity-based likelihood assigns a density to both time and mark.
Second, one can choose $\lambda_m(t\mid \E_{<t})$ to depend on a \emph{history-dependent residual} that mirrors \eqref{eq:event_rule}.
For example, define
\[
r(\U,p,t) = L(\U,t)-L(\U,t^{-}(\U;t)) - p\,C(\U),
\]
and then apply a smooth positive mapping from $r$ onto $\lambda_{(\U,p)}$ (see Section~\ref{sec:likelihood}). 
Under this construction, the point-process model can be viewed as a probabilistic relaxation of hard contrast-threshold triggering.
It avoids explicit root finding for threshold crossing times while yielding a tractable objective (negative log-likelihood) for gradient-based learning
with continuous-time latent dynamics.

% =========================
% Problem formulation (separate section)
% =========================
\section{Problem formulation}\label{sec:problem}
Let us consider an event camera observing a moving scene (e.g., a moving object or a moving camera) whose motion is governed by a latent continuous-time state $\X(t)\in\mathbb{R}^n$.
The state evolves according to continuous-time dynamics parameterized by unknown parameters $\vartheta$ \eqref{eq:neuralode}.
The camera outputs only the resulting asynchronous event stream \eqref{eq:event_stream}. 
Event timing depends on changes in log-intensity and on the contrast threshold \eqref{eq:event_rule}. 
Here, we treat the threshold as unknown and parameterize it as $C(\U)=C_{\psi}(\U)$ for any $\U \in \Omega$ with learnable parameters $\psi$.
Given $\mathcal{E}$, our goal is to estimate $(\vartheta,\psi)$ by maximum likelihood under a marked point-process observation model \eqref{eq:pp_ll_bg}.

To connect the latent state trajectory to the event measurements, we use a differentiable state-to-image model that \textit{predicts} the {intensity} $\hat L$:
\begin{equation}\label{eq:obs_model}
\hat L(\U,t;\vartheta)=\mathcal{R}\big(\U, \X (t)\big),
\end{equation}
where $\mathcal{R}$ is a renderer (or decoder) that maps the state to the image domain (note that dependence on $\vartheta$ is through $\X(t)$). Let $t^{-}(\U;t)$ denote the most recent event time at pixel $\U \in \Omega$ strictly before time $t$.
We define the history-dependent log-intensity increment relative to $t^{-}(\U;t)$ as
\begin{equation}\label{eq:dLhat_def}
\Delta \hat L(\U,t;\vartheta)=\hat L(\U,t;\vartheta)-\hat L\big(\U,t^{-}(\U;t);\vartheta\big).
\end{equation}
Under the contrast-threshold rule \eqref{eq:event_rule}, an event at pixel $u$ with polarity $p$ is triggered when the increment reaches the threshold.
Accordingly, we expect the threshold-crossing residual
$\Delta \hat L(\U,t;\vartheta)-p\,C_{\psi}(\U)$
to be close to zero at event times.
In the following section, we construct a history-dependent point-process intensity by mapping this residual to a positive event rate, yielding a differentiable maximum-likelihood objective.

\section{Proposed Method}\label{sec:method}
\subsection{Overview of the proposed method}
Event streams can be long, so we seek per-update computation that does not grow with elapsed time.
In our model, the history dependence enters only through the most recent event at the same pixel, so the event likelihood
can be evaluated using the per-pixel memory $\big(t^{-}(\U),\hat L^{-}(\U)\big)$ defined in the previous section.
The compensator term used in the log-likelihood \eqref{eq:pp_ll_bg} requires summing intensities over the full pixel grid, which we approximate by Monte Carlo
subsampling over pixels. We then optimize a fixed-lag receding-horizon objective on a receding horizon window and detach the boundary memory at the window start, yielding bounded backpropagation through the gradients across windows.

\subsection{Likelihood objective}\label{sec:likelihood}
Using \eqref{eq:dLhat_def}, let us define the residual
\begin{equation}\label{eq:residual}
\begin{aligned}
\varphi_{\U,p}(t;\vartheta,\psi)
&=\Delta \hat L(\U,t;\vartheta)-p\,C_{\psi}(\U).
\end{aligned}
\end{equation}
We model events as samples from a history-dependent marked point process with conditional intensity
$\lambda_{\U,p}(t\mid \E_{<t};\vartheta,\psi)$.
We abbreviate $\lambda_{\U,p}(t\mid \E_{<t};\vartheta,\psi)$ as $\lambda_{\U,p}(t)$
\begin{equation}\label{eq:lambda_shorthand}
\lambda_{\U,p}(t)=\lambda_{\U,p}(t\mid \E_{<t};\vartheta,\psi).
\end{equation}
We use a smooth positive parameterization that increases as the residual approaches the threshold boundary:
\begin{equation}\label{eq:intensity}
\begin{aligned}
\lambda_{\U,p}(t)
&=\lambda_{0}
+\softplus \Big(\beta-\gamma\,\abs{\varphi_{\U,p}(t;\vartheta,\psi)}\Big),
\end{aligned}
\end{equation}
where $\softplus(s)=\log\big(1+\exp(s)\big)$ is the softplus function (a smooth approximation of $\max\{0,s\}$),
$\lambda_{0}\ge 0$ is a floor rate, $\beta\in\mathbb{R}$ sets the peak level, and $\gamma>0$ controls sharpness.
Defining the distance-to-threshold $d=\abs{\varphi_{\U,p}(t;\vartheta,\psi)}$, \eqref{eq:intensity} is a decreasing
function of $d$ with peak value $\lambda_0+\softplus(\beta)$ at $d=0$.
\begin{figure}[t]
  \centering
  \includegraphics[width=0.9\linewidth]{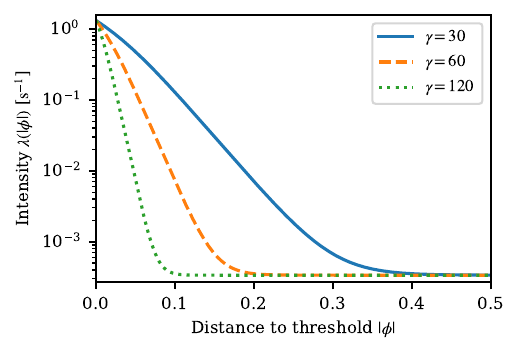}
  \caption{Surrogate intensity \eqref{eq:intensity} as a function of the distance-to-threshold $d=\abs{\varphi}$.
  Increasing $\gamma$ concentrates events closer to $d=0$ (here shown for $\gamma\in\{30,60,120\}$; log-scale on the $y$-axis).}
  \label{fig:intensity_shape}
\end{figure}
Fig.~\ref{fig:intensity_shape} illustrates how $\gamma$ controls the concentration of probability mass near $d\approx 0$.
For large $\gamma$ (and small $\lambda_{0}$), events concentrate near $\abs{\varphi_{\U,p}(t;\vartheta,\psi)}\approx 0$,
providing a differentiable surrogate of contrast-threshold triggering.
In this paper, we use \eqref{eq:intensity} as a practical likelihood model, and learn $(\vartheta,\psi)$ and treat $(\lambda_{0},\beta,\gamma)$ as fixed hyperparameters.

Let $T$ be the final observation time and define the total intensity
$\Lambda(t)=\sum_{\U\in\Omega}\sum_{p\in\{+1,-1\}}\lambda_{\U,p}(t)$.
Under a marked temporal point process, the probability density of observing the marked event stream $\E$ on $[t_0,T]$ can be written as
\begin{equation}\label{eq:pp_factored}
\begin{aligned}
p(\E\mid\vartheta,\psi)
&=\prod_{k=1}^{K}\lambda_{\U_k,p_k}(t_k)\,
\exp\!\left(-\int_{t_0}^{T}\Lambda(t)\,dt\right),\\
\Lambda(t)
&=\sum_{\U\in\Omega}\sum_{p\in\{+1,-1\}}\lambda_{\U,p}(t).
\end{aligned}
\end{equation}
Taking the logarithm yields
\begin{equation}\label{eq:pp_ll}
\begin{aligned}
\log p(\E\mid\vartheta,\psi)
&=\sum_{k=1}^{K}\log \lambda_{\U_k,p_k}(t_k)
-\int_{t_0}^{T}\Lambda(t)\,dt.
\end{aligned}
\end{equation}
Then, we minimize $\ell (\vartheta,\psi)=-\log p(\E\mid\vartheta,\psi)$.
The exponential factor in \eqref{eq:pp_factored} is often called a \textit{compensator} (or, {survival}) term: it is the probability of generating \emph{no additional events} beyond those observed, so that it utilizes information for regions where the stream is silent.
%Without the integral term, the objective would depend only on observed events and would not correspond to a valid probability density.Moreover, maximizing only the event term $\sum_k \log\lambda_{\U_k,p_k}(t_k)$ would encourage arbitrarily large intensities; the compensator $\int \Lambda(t)\,dt$ counterbalances this by penalizing excessive total event rates.
\subsubsection{Monte Carlo approximation and augmented dynamics}
Summing over all pixels (and both polarities) at every time is expensive, as it requires evaluating $\lambda_{\U,p}(t)$ over the full $2|\Omega|$ mark space repeatedly for high-resolution sensors.
We approximate the pixel sum by Monte Carlo sampling:
\begin{equation}\label{eq:mc_survival}
\sum_{\U\in\Omega}\sum_{p}\lambda_{\U,p}(t)
\approx
\frac{|\Omega|}{S}\sum_{s=1}^{S}\sum_{p\in\{+1,-1\}}
\lambda_{\tilde{\U}_s,p}(t),
\end{equation}
where $\tilde{\U}_s$ are pixels sampled uniformly from $\Omega$ and $S$ is the number of sampled pixels.
A standard implementation augments the ODE with an auxiliary state $z(t)$ to accumulate the survival integral:
\begin{equation}\label{eq:aug_ode}
\frac{d}{dt}
\begin{bmatrix}
\X(t)\\ z(t)
\end{bmatrix}
=
\begin{bmatrix}
f_{\vartheta}(\X(t),t)\\
\sum_{\U,p}\lambda_{\U,p}(t)
\end{bmatrix}.
\end{equation}
Integrating \eqref{eq:aug_ode} yields the compensator term in the marked point-process log-likelihood, which can be used for the forward update:
\begin{equation}
z(T)-z(t_0)=\int_{t_0}^{T}\Lambda(t)\,dt.
\end{equation}
\subsection{Receding-horizon estimation}\label{sec:rh}
Let $t_0 =\tau_0<\tau_1<\cdots<\tau_M=T$ be the parameter-update times. In this paper, for simplicity, this streaming schedule is given by the update period $\Delta t_{\mathrm{upd}}$, i.e., $\tau_m=\tau_0+m\,\Delta t_{\mathrm{upd}}$.
Rather than repeatedly optimizing over the growing prefix $\E_{\le \tau_m}$, we update parameters on a \emph{moving time window}.
Introduce a horizon length $\Delta>0$ and define the window event set
\begin{equation}\label{eq:window_events}
\E_{(\tau_m-\Delta,\tau_m]}=\{(\U_k,t_k,p_k)\in\E\mid \tau_m-\Delta < t_k \le \tau_m\}.
\end{equation}
\subsubsection{Windowed negative log-likelihood with detached boundary memory}
We utilize the event history up to the window start by a \emph{boundary memory}, which we denote by 
$\mathsf{s}_{\tau_m-\Delta}$.
To be specific, for each pixel $\U\in\Omega$, define the boundary last-event time
\begin{equation}\label{eq:bdry_last_time}
\begin{aligned}
t^{-}_{\tau_m-\Delta}(\U)
&= \max_{k:\,\U_k=\U,\; t_k\le \tau_m-\Delta}\; t_k,
\end{aligned}
\end{equation}
Then, we collect these two per-pixel quantities as
\begin{equation*}
\mathsf{s}_{\tau_m-\Delta} = \Big\{ \big(t^{-}_{\tau_m-\Delta}(\U),\, \hat L^{-}_{\tau_m-\Delta}(\U)\big)\Big\}_{u\in\Omega}.
\end{equation*}
Conditioned on this memory, the point-process likelihood over the window has the same form as \eqref{eq:pp_factored}--\eqref{eq:pp_ll}, but restricted to the interval $[\tau_m-\Delta,\tau_m]$. We define the windowed objective
\begin{equation}\label{eq:window_loss}
\begin{aligned}
\ell^{\mathrm{win}}_m(\vartheta,\psi)=&
-\sum_{(\U_k,t_k,p_k)\in \E_{(\tau_m-\Delta,\tau_m]}} \log \lambda_{\U_k,p_k}(t_k)
\\ & +\int_{\tau_m-\Delta}^{\tau_m}\Lambda(t)\,dt.
\end{aligned}
\end{equation}
To evaluate \eqref{eq:window_loss}, we initialize the memory at $\tau_m-\Delta$ with $\mathsf{s}_{\tau_m-\Delta}$, replay events in $\E_{(\tau_m-\Delta,\tau_m]}$ in chronological order to compute the event term, and approximate the compensator integral using the Monte Carlo approximation \eqref{eq:mc_survival}.
To keep the computation graph bounded, we \emph{detach} gradients through $\mathsf{s}_{\tau_m-\Delta}$, so backpropagation is restricted to the current window. At each update time $\tau_m$, we take a small number of gradient steps on \eqref{eq:window_loss}, which plays a similar role to a few iterations in moving-horizon estimators.

\subsection{Training and adjoint gradients for event likelihoods}\label{sec:training}
Recall that our online updates minimize the windowed negative log-likelihood \eqref{eq:window_loss}; 
the objective consists of a sum over discrete event times and a continuous-time compensator.
Since it exhibits continuous-time and discrete terms, the adjoint evolves continuously between events and receives jump updates at event times.

For update $m$, let us denote the window interval for the notational simplicity $[t_a,t_b]=[\tau_m-\Delta,\tau_m]$ and condition on the detached boundary memory
$\mathsf{s}_{\tau_m-\Delta}$ at $t_a$ (Section~\ref{sec:rh}).
We define the adjoint
\[
\mathbf{a}(t)=\frac{\partial \ell^{\mathrm{win}}_m(\vartheta,\psi)}{\partial \X(t)}, \ \   t\in[t_a,t_b].
\]

\textit{Between event times}, $\X(t)$ follows $\dot{\X}(t)=f_{\vartheta}(\X(t),t)$ and the adjoint satisfies
\begin{equation}\label{eq:adjoint_dyn}
\frac{d\mathbf{a}(t)}{dt}
=
-\Big(\frac{\partial f_{\vartheta}(\X(t),t)}{\partial \X}\Big)^{\!\top}\mathbf{a}(t)
-\frac{\partial \Lambda(t)}{\partial \X},
\end{equation}
where $\Lambda(t)=\sum_{\U\in\Omega}\sum_{p\in\{+1,-1\}}\lambda_{\U,p}(t)$ is the total intensity (approximated by the Monte Carlo estimator
\eqref{eq:mc_survival} in practice).
By the chain rule,
\begin{align}
\frac{\partial \Lambda(t)}{\partial \X}
&=
\sum_{\U\in\Omega}\sum_{p\in\{+1,-1\}}
\frac{\partial \lambda_{\U,p}(t)}{\partial \varphi_{\U,p}(t)}
\frac{\partial \varphi_{\U,p}(t)}{\partial \X}, \label{eq:Lambda_x_chain}\\
\frac{\partial \lambda_{\U,p}(t)}{\partial \varphi_{\U,p}(t)}
&=
-\gamma\,\sigma \big(\beta-\gamma|\varphi_{\U,p}(t)|\big)\,\mathrm{sign} \big(\varphi_{\U,p}(t)\big),
\label{eq:dlambda_dvarphi}
\end{align}
where $\sigma(\cdot)$ is the logistic sigmoid (since $d\,\mathrm{softplus}(z)/dz=\sigma(z)$).
Conditioned on the per-pixel memory $(t^{-}(\U),\hat L^{-}(\U))$ within the window,
$\varphi_{\U,p}(t)=\hat L(\U,t;\vartheta)-\hat L^{-}(\U)-p\,C_\psi(\U)$, hence
$\partial \varphi_{\U,p}(t)/\partial \X=\partial \hat L(\U,t;\vartheta)/\partial \X
=\partial \mathcal{R}(\U,\X(t),t)/\partial \X$ (recall that $\mathcal{R}$ is defined in \eqref{eq:obs_model}).
When using the Monte Carlo estimator \eqref{eq:mc_survival}, we simply differentiate through the sampled sum.

\textit{For each event time} $(\U_k,t_k,p_k)\in \E_{(\tau_m-\Delta,\tau_m]}$, the loss receives the discrete contribution
$-\log \lambda_{\U_k,p_k}(t_k)$, which induces an adjoint jump
\begin{align}\label{eq:adjoint_jump}
\mathbf{a}(t_k^-)
=
\mathbf{a}(t_k^+)
-\frac{\partial}{\partial \X}\log \lambda_{\U_k,p_k}(t_k).
\end{align}
For the point-process event term, the jump in \eqref{eq:adjoint_jump} uses
\begin{align*}
\frac{\partial}{\partial \X}\log \lambda_{\U_k,p_k}(t_k)
=
\frac{1}{\lambda_{\U_k,p_k}(t_k)}
\frac{\partial \lambda_{\U_k,p_k}(t_k)}{\partial \varphi}
\frac{\partial \varphi_{\U_k,p_k}(t_k)}{\partial \X},
\end{align*}
with $\partial \lambda/\partial \varphi$ given in \eqref{eq:dlambda_dvarphi} and
$\partial \varphi/\partial \X=\partial \hat L(\U_k,t_k;\vartheta)/\partial \X$ as above.

The gradient with respect to dynamics parameters is accumulated as
\begin{equation}\label{eq:adjoint_grad}
\frac{d\ell^{\mathrm{win}}_m}{d\vartheta}
=
\int_{\tau_m-\Delta}^{\tau_m}
\mathbf{a}(t)^{\top}
\frac{\partial f_{\vartheta}(\X(t),t)}{\partial \vartheta}\,dt,
\end{equation}
and the gradient with respect to $\psi$ is
\begin{align}\label{eq:grad_psi}
\frac{\partial\ell^{\mathrm{win}}_m}{\partial \psi}
= &-\sum_{(\U_k,t_k,p_k)\in \E_{(\tau_m-\Delta,\tau_m]}}
\frac{\partial}{\partial \psi}\log \lambda_{\U_k,p_k}(t_k) \notag \\ 
& +\int_{\tau_m-\Delta}^{\tau_m}\frac{\partial \Lambda(t)}{\partial \psi}\,dt.
\end{align}
Note that $\psi$ enters the likelihood only through the observation model
(via $C_{\psi}(\U)$ inside $\lambda_{\U,p}(t)$), hence $\X(t)$ is independent of $\psi$
(i.e., $\partial \X(t)/\partial \psi = 0$).

\subsection{Algorithm summary}\label{sec:algorithm}
We summarize the receding-horizon estimator that turns a continuous event stream into a sequence of parameter updates. As previously stated, the update schedule is $\tau_m=\tau_0+m\,\Delta t_{\mathrm{upd}}$ and the number of replay steps per update is denoted as $N_{\mathrm{step}}$. The horizon length $\Delta$ specifies the time window optimized at each update.
At each update time $\tau_m$, we form the window event set $\E_{(\tau_m-\Delta,\tau_m]}$ and update the parameters $(\vartheta,\psi)$ by minimizing the windowed objective \eqref{eq:window_loss} using $N_{\mathrm{step}}$ Adam steps, replaying only events in the current window.
The per-pixel memory at the window boundary is carried between the updates and detaches the computation graph, which bounds backpropagation depth. Algorithm~\ref{alg:online_pp} summarizes the above procedure.

\begin{algorithm}[t]
\caption{Receding-horizon event-based parameter estimation (windowed MLE)}
\label{alg:online_pp}
\begin{algorithmic}[1]
\Require Event stream $\E=\{(\U_k,t_k,p_k)\}_{k=1}^{K}$; update times $\{\tau_m\}_{m=0}^{M}$ (or, the update period $\Delta t_{\mathrm{upd}}$); horizon length $\Delta$; replay steps per update $N_{\mathrm{step}}$; models $f_{\vartheta}$ and $\hat L(\cdot)$; intensity \eqref{eq:intensity}; MC pixels $S$; step size $\eta$.
\Ensure Parameter estimates $\{(\vartheta_m,\psi_m)\}_{m=0}^{M}$.
\State Initialize $(\vartheta_0,\psi_0)$ and $\X(\tau_0)$; set $(\vartheta,\psi)\gets(\vartheta_0,\psi_0)$.
\State Initialize stored per-pixel memory $\mathsf{s}_0$: $\forall \U\in\Omega$, set $t^{-}(\U)\gets\tau_0$, $\hat L^{-}(\U)\gets \hat L(\U,\tau_0;\vartheta)$.
\For{$m=1$ to $M$}
  \State Form window events $\E_{(\tau_m-\Delta,\tau_m]}$.
  \For{$e=1$ to $N_{\mathrm{step}}$} %\Comment{replay steps within this window}
    \State Evaluate window loss $\ell^{\mathrm{win}}_m$ \eqref{eq:window_loss} using $\mathsf{s}_{\tau_m-\Delta}$ as the initial (detached) memory. 
    %\Statex \hspace{\algorithmicindent}Replay events in $\E_{(\tau_m-\Delta,\tau_m]}$ in time order, updating temporary memory $(t^{-},\hat L^{-})$.
    \Statex \hspace{\algorithmicindent}Approximate the compensator on $[\tau_m-\Delta,\tau_m]$ by MC pixel subsampling \eqref{eq:mc_survival}.
    \State Update parameters $(\vartheta,\psi)\gets(\vartheta,\psi)-\eta\nabla \ell^{\mathrm{win}}_m$.
  \EndFor
  \State Update the memory $s_m$ by processing events in $\mathcal{E}_{(\tau_m-\Delta,\tau_m]}$: for each $(u_k,t_k,p_k)\in \mathcal{E}_{(\tau_m-\Delta,\tau_m]}$,
set $t^{-}(u_k)\leftarrow t_k$ and $\hat L^{-}(u_k)\leftarrow \hat L(u_k,t_k;\vartheta)$.
  %Update stored memory $\mathsf{s}_m$ by processing $\E_{(\tau_m-\Delta,\tau_m]}$ once under the final $(\vartheta,\psi)$.
  \State Store $(\vartheta_m,\psi_m)\gets(\vartheta,\psi)$.
\EndFor
\end{algorithmic}
\end{algorithm}

\subsection{Computational cost and memory}\label{sec:cost}
For a horizon of duration $\Delta$, let $K_m^{\mathrm{win}}=|\E_{(\tau_m-\Delta,\tau_m]}|$ be the number of events in the current window. One evaluation of the windowed objective \eqref{eq:window_loss} costs $\mathcal{O}(K_m^{\mathrm{win}})$ for the event term plus $\mathcal{O}(S)$ intensity evaluations for the Monte Carlo compensator approximation \eqref{eq:mc_survival}.
With $N_{\mathrm{step}}$ replay steps per update, the per-update cost scales as
\begin{equation}
\mathcal{O}\big(N_{\mathrm{step}}(K_m^{\mathrm{win}}+S)\big),
\end{equation}
which is bounded when $\Delta$ is fixed.
Memory usage is dominated by the per-pixel memory $\mathcal{O}(|\Omega|)$ and storing the current window events $\mathcal{O}(K_m^{\mathrm{win}})$.

\section{Numerical Experiments}\label{sec:experiments}
We evaluate the proposed receding-horizon maximum-likelihood estimator on a synthetic event-camera sequence. All experiments are implemented in Python/PyTorch and executed on Google Colab (with a CUDA-enabled GPU runtime).
Key parameters and hyperparameters are summarized in Table~\ref{tab:exp_params}.

\subsection{Problem setting}\label{sec:spiral}
We render a moving Gaussian blob on an image grid of size $H=W=64$ and generate events by the ideal
contrast-threshold rule \eqref{eq:event_rule}.
The goal is to estimate the continuous-time dynamics parameters and a pixel-dependent contrast threshold map
from the event stream.

\subsubsection{Latent center and dynamics}
Let $\mathbf{c}(t)\in \mathbb{R}^2$ denote the blob center given by 
\begin{equation}\label{eq:center_def}
\mathbf{c}(t)=\mathbf{c}_0(t)+\mathbf{z}(t),\ \mathbf{z}(t)=[x(t),y(t)]^\top,
\end{equation}
where $\mathbf{c}_0(t)$ is a known base-center drift and is given by 
\begin{equation}\label{eq:center_drift}
\mathbf{c}_0(t)=\mathbf{c}_{\mathrm{base}}+
\begin{bmatrix}
A \sin(2\pi t/T_1)\\
B \sin(2\pi t/T_2)
\end{bmatrix},\ %\mathbf{c}_{\mathrm{base}}=(32,32), 
\end{equation}
where $\mathbf{c}_{\mathrm{base}}=(32,32)$ and $A=B=22, T_1 = 1, T_2 = 1.3$ are given constants. 
The offset state $\mathbf{z}(t)$ follows the stable-focus ODE
\begin{align}\label{eq:spiral_ode}
\dot{x}(t) &= -\alpha\,x(t) - \omega\,y(t),\\
\dot{y}(t) &= \ \omega\,x(t) - \alpha\,y(t),
\end{align}
with initial condition $[x(0),y(0)]=[12.0,\,0.0]$.
In our synthetic data, events are generated using $(\alpha,\omega)=(0.265,\,7.52)$.
\subsubsection{State-to-image observation model}
Given $\mathbf{c}(t)$, the rendered intensity is given by the Gaussian-blob model:
\begin{equation}\label{eq:spiral_intensity}
I(\U,t)=I_{\mathrm{bg}}+I_{\mathrm{amp}}\exp\!\left(-\frac{\|\U-\mathbf{c}(t)\|^2}{2\sigma^2}\right),
\end{equation}
with fixed $(I_{\mathrm{bg}},I_{\mathrm{amp}},\sigma)$ (Table~\ref{tab:exp_params}),
where $\U=(u_x,u_y)\in\Omega$ is a pixel location (in pixel units), consistent with the event-stream notation in Section~II-B.
The parameters $I_{\mathrm{bg}}$ and $I_{\mathrm{amp}}$ set the background level and the contrast (amplitude) of the rendered blob.
We do not estimate them from events: event generation depends on \emph{log-intensity increments} $\Delta L(\U,t)$, so additive offsets in $L$ do not change events, and the overall brightness scale is weakly identifiable and strongly coupled to the unknown contrast threshold.
Fixing $(I_{\mathrm{bg}}, I_{\mathrm{amp}})$ removes this ambiguity and isolates dynamics and threshold estimation.
The parameter $\sigma$ is the spatial standard deviation of the Gaussian blob (in pixels) and thus controls the apparent object size; larger $\sigma$ activates more pixels and increases event density, while smaller $\sigma$ produces sparser events. We define the log-intensity used for event synthesis as 
\begin{align}
    \hat L(\U,t;\vartheta)
&= \mathcal{R}\big(\U,\mathbf{c}(t)\big)
= \log\!\big( I(\U,t)+\epsilon\big)
\end{align}
 where the small parameter $\epsilon=10^{-3}$ is added for numerical stability.
\subsubsection{Event synthesis and pixel-dependent threshold}
\begin{figure}[t]
\centering
\includegraphics[width=0.8\columnwidth]{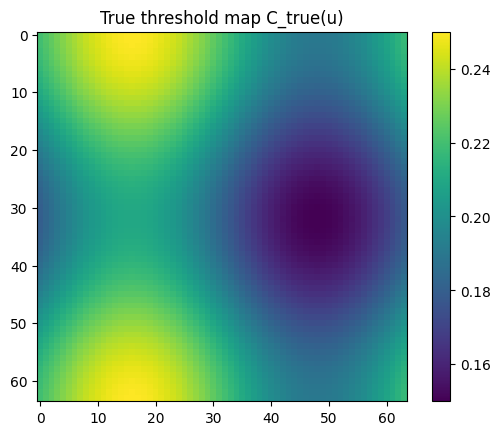}
\caption{True pixel dependent threshold map $C(\cdot)$.}
\label{fig:threshold}
\end{figure}

\begin{figure*}[t]
  \centering
  \includegraphics[width=\textwidth]{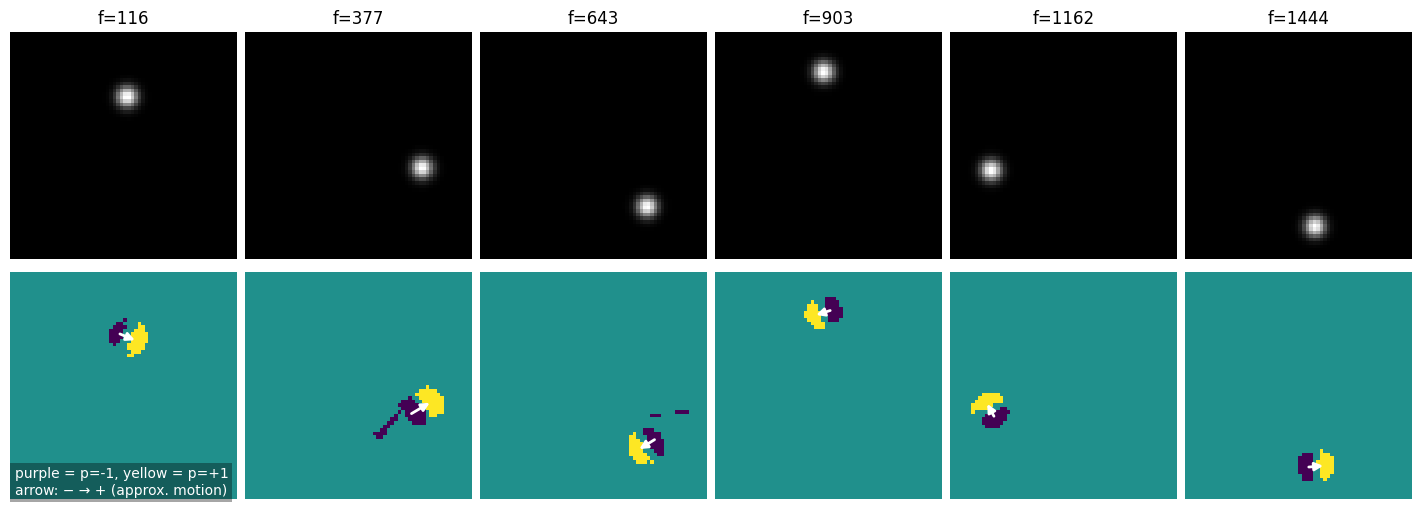}
  \caption{Representative snapshots from the synthetic event-camera sequence.
  Top: rendered intensity frames. Bottom: corresponding event images in polarity mode (yellow: $p=+1$, purple: $p=-1$, white arrow: approximated motion).}
  \label{fig:snapshots}
\end{figure*}

We render intensity frames at $\mathrm{fps}=120$ and generate events by threshold crossings of \eqref{eq:event_rule}. The sequence duration is $T = 13$s. We consider a pixel-dependent threshold field
\begin{equation}\label{eq:Ctrue_field}
C(\U)=
C_{\mathrm{base}} + 0.03 \sin\!\left(2\pi \frac{u_x}{W_{\mathrm{img}}}\right)
      + 0.02 \cos\!\left(2\pi \frac{u_y}{H_{\mathrm{img}}}\right),
\end{equation}
with $C_{\mathrm{base}} = 0.2$, where $u_x$ and $u_y$ represents respectively the $x$ and the $y$ of the pixel location $\U$.
Fig.~\ref{fig:threshold} shows the true threshold map $C(\U)$ and
Fig.~\ref{fig:snapshots} shows representative frame and event snapshots.

\begin{table}[t]
\centering
\caption{Estimated parameters and key hyperparameters in the numerical experiments (pixel-dependent threshold only).}
\label{tab:exp_params}
\small
\begin{tabular}{p{0.50\linewidth}p{0.42\linewidth}}
\hline
\textbf{Item} & \textbf{Value / description} \\
\hline
\multicolumn{2}{l}{\emph{Renderer / data (fixed)}}\\
Grid size $(H_{\mathrm{img}}, W _{\mathrm{img}})$ & $64\times 64$ \\
Base center & $c_{\mathrm{base}}=(32,32)$ \\
Drift parameters & $A=B=22$ px, $T_1=1.0$s, $T_2=1.3$s \\
Intensity params & $(I_{\mathrm{bg}},I_{\mathrm{amp}},\sigma)=(0.15,\,0.75,\,2.0)$ \\
Log eps (data / model) & $\epsilon=10^{-3}$ \\
Frame rate \& duration & $\mathrm{fps}=120$, $T=13.0$s \\
\hline
\multicolumn{2}{l}{\emph{Ground truth (synthetic)}}\\
Dynamics (used to generate data) & $(\alpha,\omega)=(0.265,\,7.52)$ \\
True threshold field & Eq.~\eqref{eq:Ctrue_field}\\
\hline
\multicolumn{2}{l}{\emph{Unknown parameters to be estimated}}\\
Dynamics & $\vartheta=\{\alpha,\omega\}$ \\
Threshold parameters & $\psi=\{C_{\mathrm{base}},\eta\}$ in Eq.~\eqref{eq:C_param} \\
\hline
\multicolumn{2}{l}{\emph{Likelihood approximation}}\\
Update cadence; replay steps & $\Delta t_{\mathrm{upd}}=0.4$s; $N_{\mathrm{step}}=30$ \\
MC pixels & $S=512$ \\
Fixed-lag horizon & $\Delta=H\,\Delta t_{\mathrm{upd}}$ \\
\hline
\end{tabular}
\end{table}

\subsection{Estimation task}\label{sec:exp_setup}
For the dynamics parameters, we aim to estimate
\begin{equation}
\vartheta=\{\alpha,\omega\}. 
\end{equation}
%and the threshold parameters $\psi$ that define a pixel-dependent contrast threshold map $C_\psi(u)$used in the point-process intensity model.
Directly learning all the thresholds within $\Omega$ requires $H_{\mathrm{img}}W_{\mathrm{img}}$ parameters
($H_{\mathrm{img}}=W_{\mathrm{img}}=64$, hence $H_{\mathrm{img}}W_{\mathrm{img}}=4096$) and is poorly conditioned in regions with few/no events.
Thus, we use a low-dimensional parameterization:
\begin{equation}\label{eq:C_param}
\psi=\{C_{\mathrm{base}},\eta\},\qquad
C_\psi(\U)=C_{\mathrm{base}}+\delta_\eta(\U),
\end{equation}
where $C_{\mathrm{base}}$ is a global offset and $\eta\in\mathbb{R}^{H^c _{\mathrm{img}}\times W^c _{\mathrm{img}}}$ is a learnable coarse-grid field
with $H^c _{\mathrm{img}}\times W^c _{\mathrm{img}}=8\times 8$.
Then, for each pixel $\U\in\Omega$, the residual $\delta_\eta(\U)$ is obtained by converting the coarse-grid field
$\eta\in\mathbb{R}^{H^c _{\mathrm{img}}\times W^c _{\mathrm{img}}}$ to a full-resolution map on the $H_{\mathrm{img}}\times W _{\mathrm{img}}$ pixel grid via bilinear interpolation.
Concretely, $\delta_\eta(\U)$ is computed as a convex combination (weighted average) of the four coarse-grid values
surrounding the location of $u$, with nonnegative weights that sum to one. With this choice, the number of unknown thresholds is $1+H^c _{\mathrm{img}}W^c _{\mathrm{img}}=65$ (one scalar $C_{\mathrm{base}}$ plus $8\times 8$ coefficients). The compensator integral is approximated by the Monte Carlo pixel subsampling
with $S=512$.
We perform receding-horizon updates every $\Delta t_{\mathrm{upd}}=0.4$s and run $N_{\mathrm{step}}=30$ replayed Adam steps per update on the current window.

\begin{figure}[t]
\centering
\includegraphics[width=1\columnwidth]{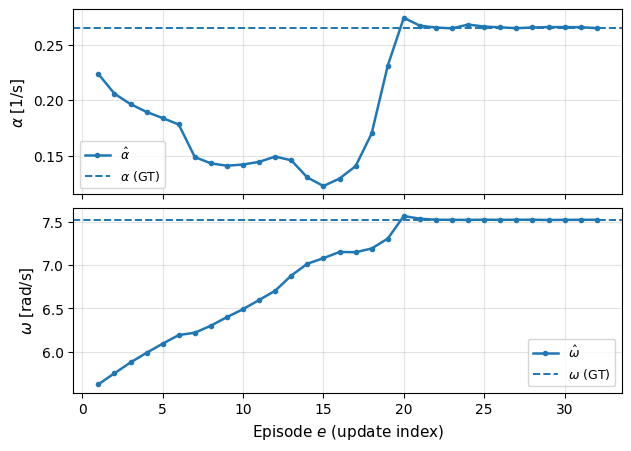}
\caption{Learning curves for a fixed horizon $H=15$.
\textbf{Top:} $\hat\alpha$ versus episode $e$.
\textbf{Bottom:} $\hat\omega$ versus episode $e$.
Dashed lines indicate the ground truth.}
\label{fig:learning_curves_H10}
\end{figure}

\begin{comment}
\begin{figure}[t]
  \centering
  \begin{subfigure}[t]{0.48\linewidth}
    \centering
    \includegraphics[width=\linewidth]{thresholds_image.png}
    \caption{True $C(\cdot)$.}
    \label{fig:threshold:true}
  \end{subfigure}\hfill%
  \begin{subfigure}[t]{0.48\linewidth}
    \centering
    \includegraphics[width=\linewidth]{thresholds_image_estimated.png}
    \caption{Estimated $\hat{C}(\cdot)$.}
    \label{fig:threshold:est}
  \end{subfigure}
  \caption{Estimted Pixel-dependent threshold maps: true $C(\cdot)$ (left) and estimated $\hat{C}(\cdot)$ (right).}
  \label{fig:threshold_result}
\end{figure}
\end{comment}

\begin{figure*}[t]
\centering
\includegraphics[width=\textwidth]{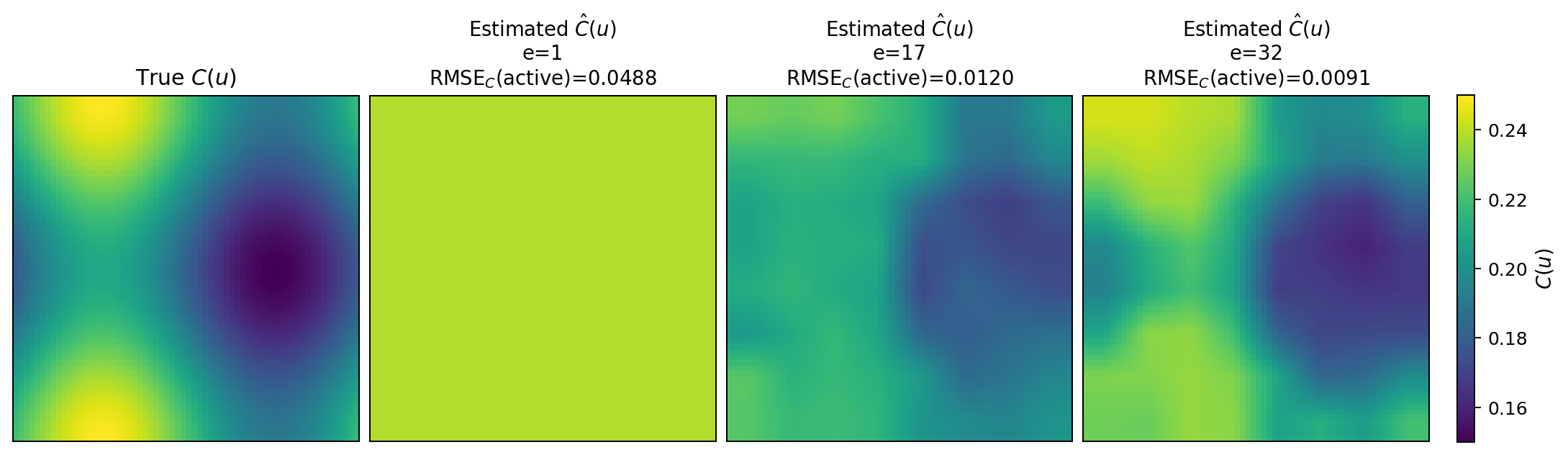}
\caption{Snapshots of pixel-dependent contrast-threshold estimation over online episodes ($H=15$).
\textbf{Left:} ground-truth threshold field $C(u)$ used for event synthesis.
\textbf{Right:} estimated maps $\hat{C}(u)$ after episodes $e\in\{1,17,32\}$.}
\label{fig:cestsnapshot}
\end{figure*}

\subsection{Simulation results}\label{sec:sim_results}

\subsubsection{Learning curves with a fixed horizon ($H=15$)}
We report simulation results on the synthetic dataset in Section~\ref{sec:spiral}.
We fix the horizon length to $H=15$ (window length $\Delta = H\,\Delta t_{\mathrm{upd}}$) and record parameter estimates after each online update.

Fig.~\ref{fig:learning_curves_H10} shows the learning curves of the dynamics parameters $(\hat\alpha,\hat\omega)$.
Both estimates exhibit a transient phase in early episodes and then converge close to the ground truth after a moderate number of updates;
after convergence, the estimates remain stable over subsequent episodes. 
Fig.~\ref{fig:cestsnapshot} visualizes how the pixel-dependent contrast threshold map is refined during receding-horizon updates.
At the first episode ($e=1$), the estimate is spatially uniform. Then, as more events are accumulated across the image, the estimator progressively recovers the spatial structure of the true field.
The remaining discrepancies are typically concentrated in regions with sparse event (i.e., regions where only a few/no events occur) activity, where the threshold is weakly constrained by the data.

\subsubsection{Horizon ablation ($H=1,\dots,34$)}
We next study the effect of the horizon length by sweeping $H\in\{1,\dots,19\}$ while keeping all other settings fixed.
We evaluate the final estimation accuracy at the end of the sequence using:
\begin{align}
\mathrm{RMSE}_\alpha(H) &= \big|\hat\alpha(H) - \alpha\big|,\\
\mathrm{RMSE}_\omega(H) &= \big|\hat\omega(H) - \omega\big|,\\
\mathrm{RMSE}_C(H) &=
\sqrt{\frac{1}{|\mathcal{A}|}\sum_{u\in\mathcal{A}}\big(\hat C_H(\U)-C(\U)\big)^2},
\label{eq:rmse_C_masked}
\end{align}
where $\mathcal{A}\subseteq\Omega$ denotes the set of \emph{active} pixels (pixels that emitted at least one event in the sequence),
so that the threshold error is not dominated by unobserved regions.

\begin{figure}[t]
\centering
\includegraphics[width=1.0\columnwidth]{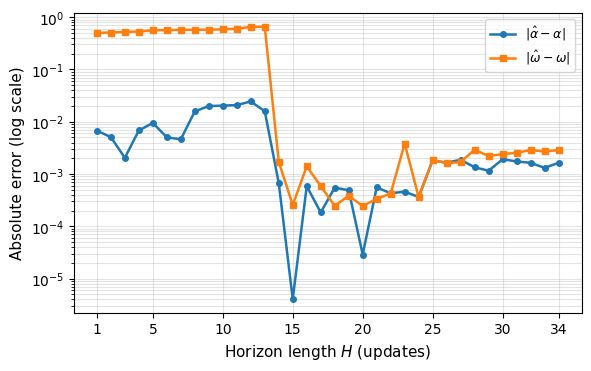}
\caption{Horizon ablation. $\mathrm{RMSE}_\alpha(H)$ and $\mathrm{RMSE}_\omega(H)$ as functions of the fixed-lag horizon $H$ (log-scale y-axis).}
\label{fig:ablation_rmse_ao}
\end{figure}

\begin{figure}[t]
\centering
\includegraphics[width=1.0\columnwidth]{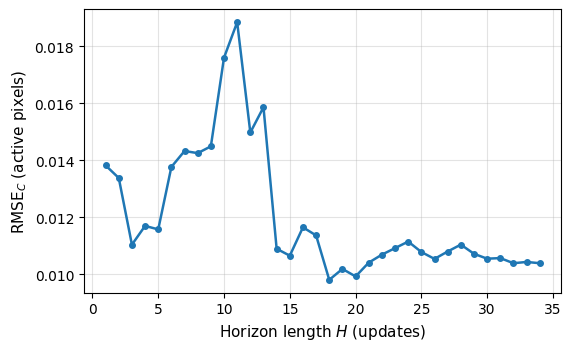}
\caption{Horizon ablation. Threshold-map error $\mathrm{RMSE}_C(H)$ over active pixels \eqref{eq:rmse_C_masked} as a function of $H$. Note that the y-axis is shown in linear scale.}
\label{fig:ablation_rmse_C}
\end{figure}

\begin{figure}[t]
\centering
\includegraphics[width=1.0\columnwidth]{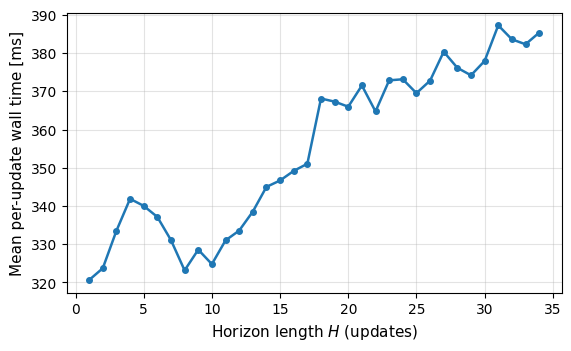}
\caption{Mean update time per step as a function of $H$.}
\label{fig:updatetime}
\end{figure}

Figs.~\ref{fig:ablation_rmse_ao} and \ref{fig:ablation_rmse_C} report the RMSE values as functions of $H$.
The estimation of $\omega$ is particularly sensitive: its error remains large for short horizons ($H\le 13$),
but drops by orders of magnitude once the horizon becomes sufficiently long ($H\ge 14$).
The RMSE of $\alpha$ shows a similar improvement in the long-horizon regime.
The threshold-map error $\mathrm{RMSE}_C$ in Fig.~\ref{fig:ablation_rmse_C} is comparatively similar across $H$ (i.e., RMSE with the order of $10^{-2}$), but becomes slightly smaller for larger horizons,
reaching its minimum around $H\approx 18$.
Fig.~\ref{fig:updatetime} shows the computational cost as a function of $H$; the mean update time increases gradually with $H$, although we note that it is still smaller than the update time interval (i.e., $0.4$s$=400$ms). 

\section{Conclusion and Future Work}
We presented an online maximum-likelihood framework for identifying continuous-time dynamics from raw event-camera streams. By combining a Neural ODE state model with a history-dependent marked point-process observation model, the proposed formulation captures the asynchronous thresholded nature of event generation while preserving a normalized likelihood over event times and marks. A differentiable surrogate of contrast-threshold triggering enables joint estimation of the dynamics and pixel-dependent thresholds, and the fixed-lag receding-horizon update with compact boundary memory and Monte Carlo compensator approximation keeps the per-update computation bounded. On synthetic data, the method recovers both the dynamics parameters and the threshold map and reveals the accuracy--latency trade-off induced by the horizon length. These results suggest that likelihood-based continuous-time identification from raw event timing is a viable alternative to event aggregation when online adaptation is required.

Future work includes validation on real event-camera data, incorporation of richer sensor nonidealities such as polarity-dependent thresholds and temporal drift, and improvement of the compensator approximation via adaptive sampling. Extending the framework to joint online state-and-parameter estimation in more complex scenes and studying uncertainty and identifiability are also important directions for future work.

\section{Acknowledgements}
This work was supported by JST, CRONOS, Japan Grant Number JPMJCS25K3.

%On a synthetic spiral sequence, the proposed method recovers the dynamics parameters $(\alpha,\omega)$ and the pixel-dependent threshold map $C(u)$.The learning curves with $H=10$ show that both $\hat{\alpha}$ and $\hat{\omega}$ converge toward the ground truth over online updates, while the estimated threshold map captures the overall spatial structure of $C(u)$ (with larger residual errors in weakly excited regions).In the horizon ablation, estimation accuracy improves substantially as the horizon grows, and we observe a clear transition in the high-lag regime: once $H$ reaches approximately $9$ or more, the RMSE drops markedly (especially for $\omega$), indicating reduced fixed-lag bias.In our setup with update cadence $\Delta t_{\mathrm{upd}}=0.2$\,s, this corresponds to a window length of about $\Delta = H\Delta t_{\mathrm{upd}} \approx 1.8$--$2.0$\,s, providing a practical accuracy--latency trade-off with bounded per-update computation. Future work includes validation on real event datasets (including timestamp quantization and sensor non-idealities), scaling to higher resolutions and richer state-to-image models, and exploring alternative threshold and intensity parameterizations (and variance-reduction strategies for the Monte Carlo compensator) to further improve robustness and efficiency in online settings.
\bibliographystyle{IEEEtran}
\bibliography{IEEE_patchedver2_compact_no_doi_with_sp_recent_v2_corrected}

\end{document}